# A Chiral Inverse Faraday Effect Mediated by an Inversely Designed Plasmonic Antenna

Ye Mou, Xingyu Yang, Bruno Gallas, and Mathieu Mivelle*

Sorbonne Université, CNRS, Institut des NanoSciences de Paris, INSP, F-75005 Paris, France

*Corresponding author: mathieu.mivelle@sorbonne-universite.fr

**Abstract**

The inverse Faraday effect is a magneto-optical process allowing the magnetization of matter by an optical excitation carrying a non-zero spin or orbital moment of light. This phenomenon was considered until now as symmetric; right or left circular polarizations generate magnetic fields oriented in the direction of light propagation or in the counter-propagating direction. Here, we demonstrate that by manipulating the spin density of light in a plasmonic nanostructure, we generate a chiral inverse Faraday effect, creating a strong magnetic field of 500 mT only for one helicity of the light, the opposite helicity producing this effect only for the mirror structure. This new optical concept opens the way to the generation of magnetic fields with unpolarized light, finding application in the ultrafast manipulation of magnetic domains and processes, such as spin precession, spin currents and waves, magnetic skyrmion or magnetic circular dichroism, with direct applications in data storage and data processing technologies.

# 1. Introduction

The inverse Faraday effect (IFE) is a magneto-optical process enabling the magnetization of matter by optical excitation only[1-3] (Figure 1). This magnetization is made possible by the action of non-linear optical forces on the matter's electrons.[4-7] In particular, in a metal, the free electrons subject to these non-linear forces are set in a drift motion at the origin of the IFE. If we consider these electrons as free-moving charges, the expression of these drift currents (**J**$_d$) can be described using the plasma community's formalism[8, 9] developed. In this case, R. Hertel demonstrated that in metal, **J**$_d$ is written as:[7, 10]

$$J_d = \frac{1}{2en} Re\left(\left(-\frac{\nabla \cdot (\sigma_\omega E)}{i\omega}\right) \cdot (\sigma_\omega E)^*\right) \quad (1)$$

With e the charge of the electron (e < 0), n the charge density at rest, $\sigma_\omega$ the dynamic conductivity of the metal, and **E** the optical electric field.

These drift currents are, therefore, a function of the optical electric field and its divergence. Because of their ability to manipulate fields and field gradients, nanophotonics and nanoplasmonics are then particularly well suited to generate a strong IFE and thus create strong stationary magnetic fields (**B**).[11-16] Moreover, a plasmonic IFE will confine the created **B**-field to sub-wavelength scales due to its nanometric scales.[11, 16-19] Finally, the generation of the **B**-field by IFE is due to light-matter interactions; therefore, by using ultra-short optical pulses, nanoplasmonics is today the only technique allowing the creation of intense, confined, and ultra-fast magnetic field pulses.[14, 16]. These unique properties have applications in many fields of magnetic research and technology.[20] Indeed, since the pioneering work of Beaurepaire *et al*,[21] researchers have been looking for ways to manipulate and study magnetization at very short time scales, mainly through the use of femtosecond lasers, intending to control and accelerate current data storage technologies. Unfortunately, the physical processes involved in this type of interaction are still poorly understood. Likewise, the transient processes of magnetic interactions, such as spin precession, spin-orbit coupling, and exchange interactions, have their roots in the femtosecond time scale.[22] The ability to probe and address these different processes and their transient mechanisms using ultrashort pulses of magnetic fields would benefit countless research activities in magnetism: from Zeeman splitting,[23] magnetic trappings,[24] magnetic skyrmions,[25] magneto-plasmonics,[26] ultrafast magnetic modulations,[27] and magnetic circular dichroism[28] to spin control,[29] spin precession,[30] spin currents,[31] and spin waves.[32]

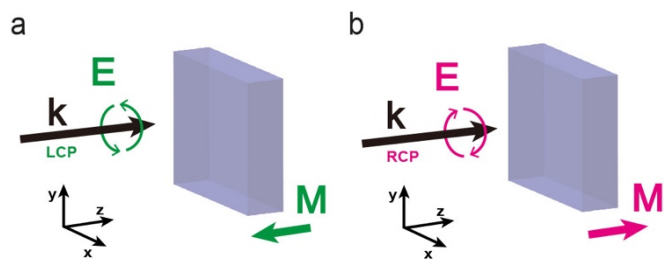

*Figure 1. Principle of the inverse Faraday effect. A circularly polarized electromagnetic wave magnetizes a material. This magnetization is oriented a) against the direction of the wave propagation when the light is left circularly polarized and b) in the direction of the light propagation when it is right circularly polarized.*

The IFE is a symmetrical process, i.e., the magnetization of matter by a right circular polarized wave will be opposite to a left circular polarized wave. In particular, the magnetic field created for a right circular polarization will be oriented in the propagation direction of the light, and that of a left circular polarization will be oriented in a counter-propagative way (Figure 1).

Here, we demonstrate the generation of a chiral IFE through the manipulation of light at the nanoscale and, in particular, via the local manipulation of the spin density (characterizing the degree of circular polarization of a wave). Using an inverse design algorithm, we have generated a plasmonic nanostructure that creates a non-zero **B**-field for a single helicity of light. Moreover, we show that the mirror image of the optimized plasmonic nanostructure generates a non-zero **B**-field oriented in the opposite direction and only for the other helicity of light. Also, under the illumination conditions used here, the amplitude of the created **B**-field is estimated to be 0.5 T, making it one of the strongest generated by a plasmonic nanostructure.[11, 12, 14, 16] Finally, this chiral IFE results from the photonic nanostructures' ability to manipulate light in the near field. Indeed, we demonstrate that this effect is due to generating a spin-density hot spot within this nanostructure for a single excitation polarization of light. These results are particularly important since they imply that this approach allows optically generating, even with unpolarized incoherent light, magnetic fields that are intense, ultrafast, nanoscale, and always oriented in the same direction. Thus, allowing for manipulation, at very short time and space scales, of many magnetic processes.

## 2. Results and discussions

Specifically, we have optimized, using a genetic algorithm (GA),[33-35] a plasmonic nanostructure made in a thin gold layer of 30 nm thickness deposited on a glass substrate (Figure 2). This nanostructure is based on a 2D matrix of 10x10 elements matrix, each element consisting of metal or air with a size of 28 nm, constituting a total structure size of 280x280 nm$^2$. These dimensions are chosen to enable the experimental fabrication of this structure by lithography techniques, for instance. For the same reasons and to avoid the non-physical effects that a numerical approach can generate locally, the corners of the nanostructure are smoothed (Figure S1). The excitation of the structure is done by a circularly polarized plane wave, with a wavelength of 800 nm, launched from the substrate side and a power of $10^{12}$ W/cm$^2$. This excitation power is chosen to be below the threshold of what the material can withstand.[36, 37] Each generation of the GA optimization is composed of 200 elements. For each element, two simulations are performed, one with a right and another with a left circular polarization. The associated drift currents are calculated from the optical response of each element under these two polarizations (Equation (1)). Hence, Biot and Savart's law estimates the field B generated under these two excitation conditions at the center of the nanostructure (symbolized by the green star in Figure 2). We then choose to maximize the difference $B_{RCP}$-abs($B_{LCP}$) as a GA optimization function, with **B**$_{RCP}$ and **B**$_{LCP}$ the **B**-fields created by a right or left circular polarization, respectively. The evolution from a generation N to N+1, N being the number of the generation, is then done by keeping the 200 best structures of the generations 1 to N. The breeding of these 200 structures produces half of the elements of the generation N+1, the other half being constituted of mutated elements with a mutation rate of 10%.

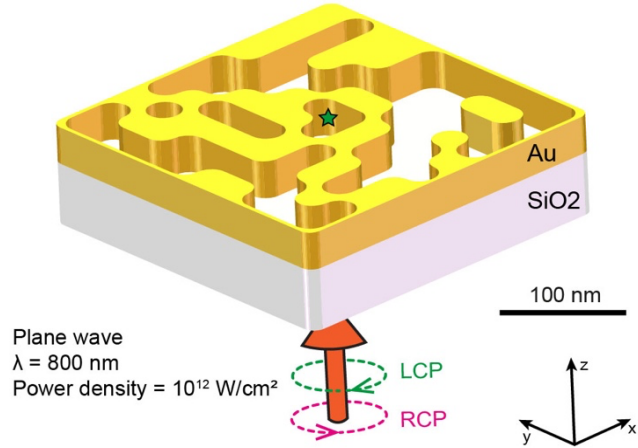

*Figure 2*. Optimized structure and excitation conditions. Example of a GA-optimized structure, realized in a 30 nm thick gold layer, for excitation by a right- or left-circular polarized plane wave at the wavelength of 800 nm and for an excitation power of $10^{12}$ W/cm$^2$.

These optical and selection characteristics give an optimized structure after 76 generations (Figure 2,3A and S2). From the optical response of this optimized structure (Figure S3) and via Equation (1), the associated drift currents are calculated inside the metal (Figure S4). Using Biot and Savart's law, we calculate and show in Figure 3B and C the Z-oriented stationary magnetic field distribution in an XY plane at the surface of the optimized structure for left and right circular polarization, respectively. As can be seen, only the right circular polarization generates an intense **B**-field of 0.5 T, oriented in the direction of light propagation (along the positive Z), with an abs($B_{RCP}/B_{LCP}$) ratio of 11. Also, using the plasmonic mirror structure to the optimized one (Figure 3D), we observe that under the same excitation conditions but for a left circular polarization, a **B**-field of the same intensity is generated but oriented in a counter propagative way to the incident light (along the negative Z) and for a left circular polarization. This is the first time a chiral IFE has been observed, moreover with this intensity and at the nanoscale.

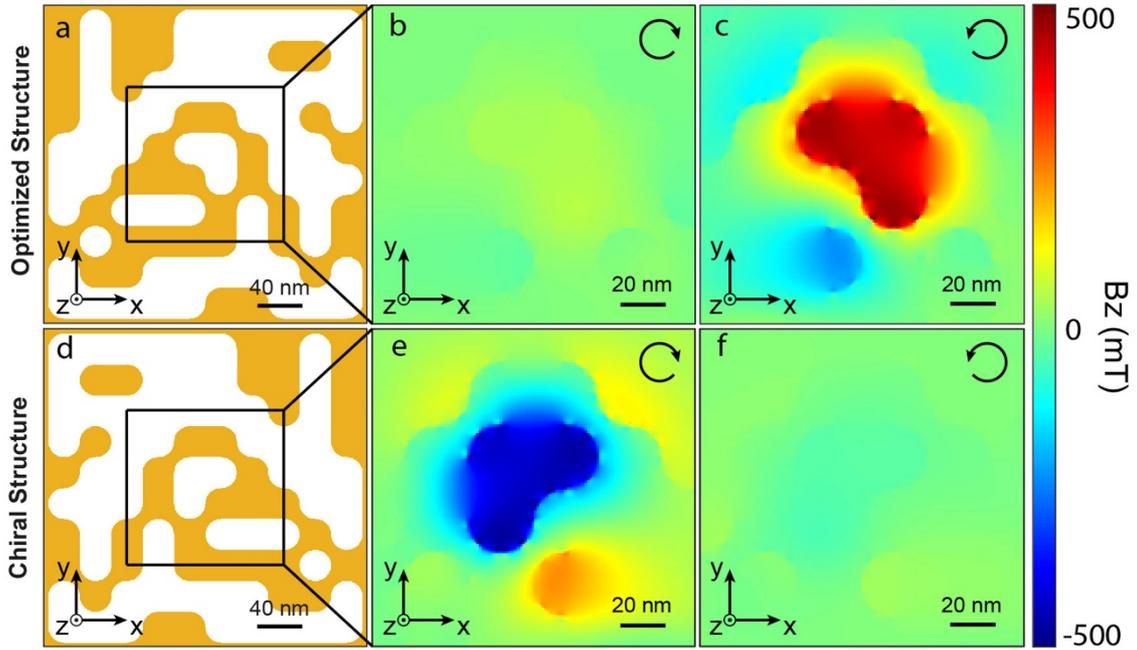

***Figure 3***. *Magnetic response of the optimized plasmonic nanostructure. a) Schematic, in an XY plane, of the GA-optimized structure. b) and c) Spatial distribution of the magnetic field oriented along Z and generated in the Z-center of the structure shown in a) for the left and right circular polarizations of excitation, respectively. d) Schematic, in an XY plane, of the mirror structure displayed in a). e) and f) Spatial distributions of the **B**-field oriented along Z and generated in the Z-center of the mirror structure shown in d) for the left and right circular polarizations of excitation, respectively. See **Figure** S5 for other **B**-field components and **Figure** S6 for its full amplitude.*

The observation of this new physical effect is due to the ability of optical nanostructures to manipulate light and its characteristics in the near field. Indeed, it is known that the manipulation of electromagnetic fields in the near field allows, for instance, to control of local densities of states,[38-40] radiation patterns,[41] chirality densities,[42, 43] or even some nonlinear effects.[44] Here, we use the unique properties of plasmonic nanostructures to manipulate the spin densities of light locally, or in other words, the local helicity of light. The equation describing the electric spin density of light is:

$$s = \frac{1}{|E_0|^2} Im(E^* \times E) \qquad (2)$$

With $E_0$ the electric field of the incoming light.

The spin density can thus take positive or negative values corresponding to right or left elliptical polarizations, respectively, 0 describing a linear polarization. In the far field, the spin density can only take values between -1 and 1, -1 being a left circular polarization and 1 a right circular polarization (Figure 4A and B). On the other hand, in the near field, the spin density normalized to the incident intensity $|E_0|^2$ can take much larger values due to the exaltation of the fields, leading to the concept of super-circular light by analogy with super-chiral light.[45] Therefore, since the generation of drift currents requires an elliptical or circular polarization (Equation (1)), by creating locally non-zero spin densities, the generation of an IFE is possible in a plasmonic structure. Here, the chiral property of our optimized design comes from the fact that for excitation for two different circular polarizations, only one polarization generates locally a non-zero spin density (Figure 4C and D), the one that is right circular (Figure 4D). Similarly, for the mirror plasmonic structure (Figure 4E and F), only one opposite light helicity generates a non-zero spin density, the left circular polarization (Figure 4E). Moreover, as can be seen in Figure 4D and E, the signs of the local spin densities generated in the nanostructures are opposite. It is positive in the case of a right circular excitation (Figure 4D), corresponding to a right elliptical polarization, and negative in the case of a left circular excitation (Figure 4E), corresponding to a left elliptical polarization. As a result of the IFE, the optimized structure thus generates a **B**-field oriented in the propagating wave direction and the mirror structure in the counter-propagating one. Finally, the high amplitude of the generated **B**-field is directly related to the super-circular nature of the light generated by these plasmonic structures, creating, in turn, strong drift currents.

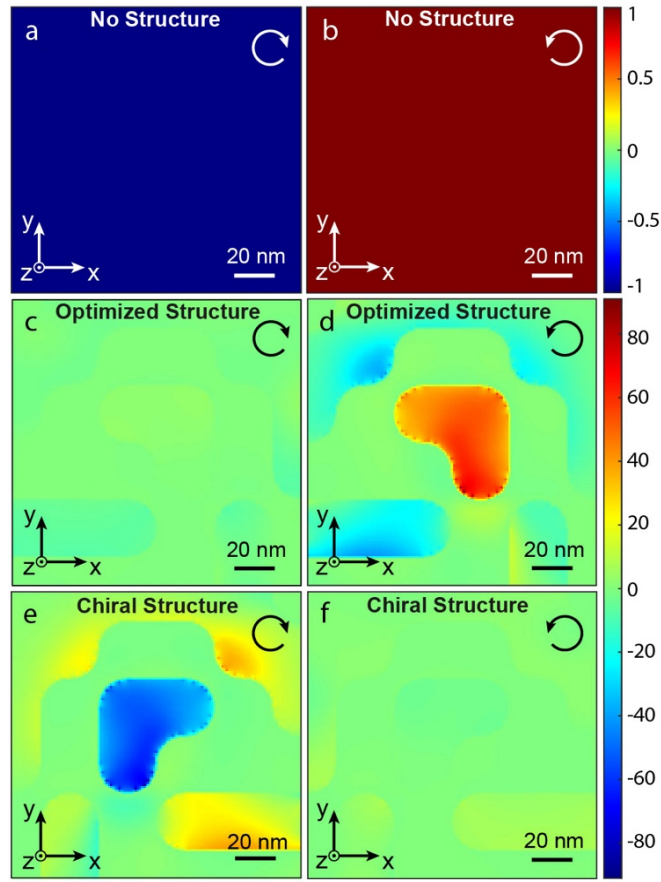

*Figure 4. Distribution of spin densities. a) And b) spin density for left and right circularly polarized plane waves, respectively. c) and d) Local spin density in the Z-center of the optimized plasmonic nanostructure for left and right circular polarization of excitation, respectively. e) and f) Local spin density in the Z-center of the mirror plasmonic structure of the optimized one, for excitation by e) a left or f) right circularly polarized wave.*

## 3. Conclusion

In conclusion, we have demonstrated for the first time that the magneto-optical process of the inverse Faraday effect can be a chiral mechanism occurring only for one helicity of light. This new physical effect is due to manipulating the light polarization at the nanoscale. Using an inverse design algorithm based on natural selection, we have shown the generation of a non-zero spin density locally in a plasmonic nanostructure for a single helicity of light, making the selective magnetization possible as a function of the excitation polarization. Also, we have demonstrated that using a mirror structure allows the generation of a **B**-field equivalent in size and amplitude but of opposite orientation, demonstrating a perfect chirality effect. Finally, due to the super-circular light created by these structures, the generated **B**-field has an amplitude of 0.5 T, which makes it one of the most intense generated at these scales and by IFE. The results presented here are significant for several reasons. The IFE by plasmonic nanostructures is the only approach allowing for the generation of ultrafast magnetic field pulses at the nanometer scale. The possibility of generating a structure-defined magnetic field for unpolarized excitation would have applications in the manipulation of magnetic processes such as skyrmion manipulation, ultrafast magnetic modulation, magnetic trapping, spin currents, or spin precession, with, for example, ultrafast data writing as a direct application.

**Author contributions:** All the authors have accepted responsibility for the entire content of this submitted manuscript and approved submission.

**Research funding:** This work is supported by the Agence National de la Recherche (ANR-20-CE09-0031-01), the Institut de Physique du CNRS (Tremplin@INP 2020) and the China Scholarship Council.

**Conflict of interest statement:** The authors declare no conflict of interest regarding this article.

# A Chiral Inverse Faraday Effect Mediated by an Inversely Designed Plasmonic Antenna

Ye Mou, Xingyu Yang, Bruno Gallas, and Mathieu Mivelle*

Sorbonne Université, CNRS, Institut des NanoSciences de Paris, INSP, F-75005 Paris, France

*Corresponding author: mathieu.mivelle@sorbonne-universite.fr

A list of the main content:

Supplementary figures S1 to S6

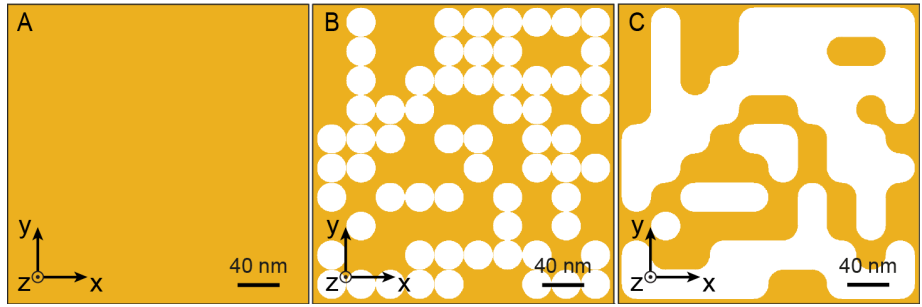

**Figure S1:** Construction of the elements constituting each generation of the genetic algorithm. Inside A) a uniform gold layer of 30 nm thickness, holes are made according to a binary matrix playing the role of the DNA in the evolutionary process. C) The obtained structure is then smoothed to avoid all the roughnesses not experimentally feasible and generating non-physical effects.

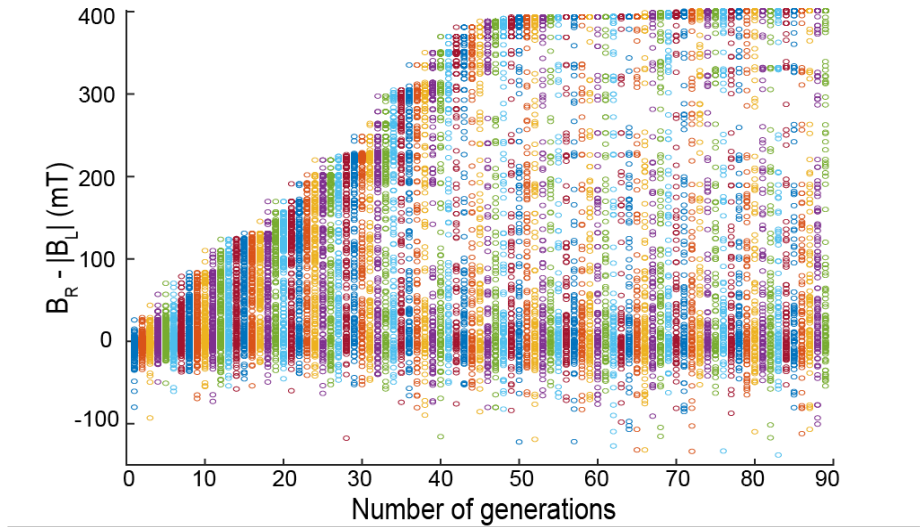

**Figure S2:** Evolutionary process. Evolution during the different generations of the optimization function consisting in maximizing the difference $B_{RCP}$-abs($B_{LCP}$), with $B_{RCP}$ and $B_{LCP}$ the B-fields created by a right or left circular polarization, respectively. Each generation consists of 200 elements. The optimized structure appears after 76 generations.

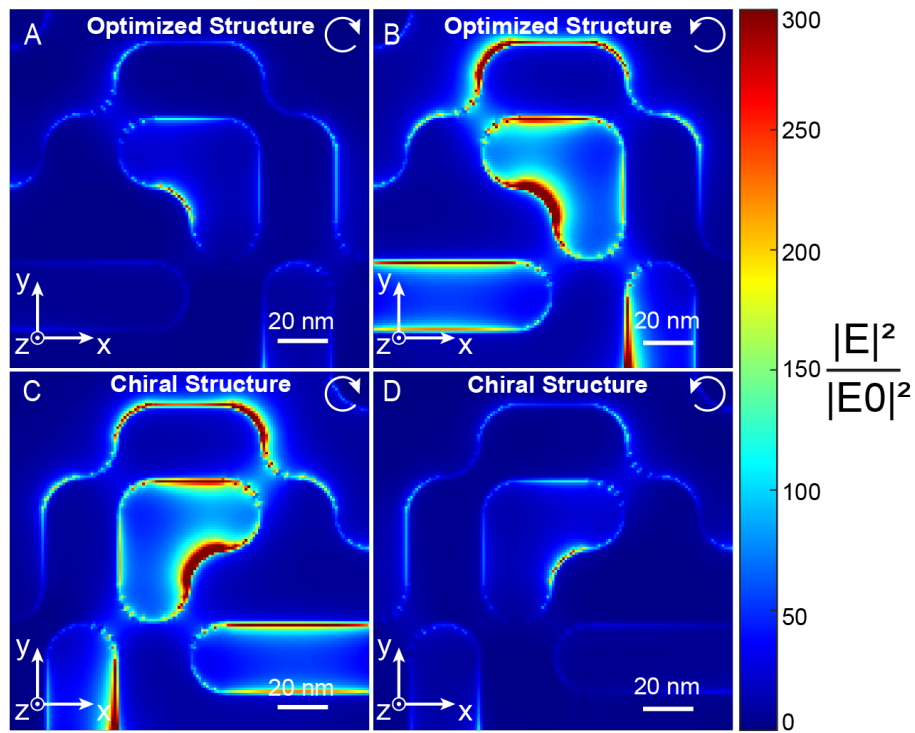

**Figure S3:** Optical responses in an XY plane. A) and B) Spatial distributions of the optical electric intensity enhancement at the surface of the optimized structure for the left and right circular polarizations of excitation, respectively. C) and D) Spatial distributions of the optical electric intensity enhancement at the surface of the mirror structure for the left and right circular polarizations of excitation, respectively.

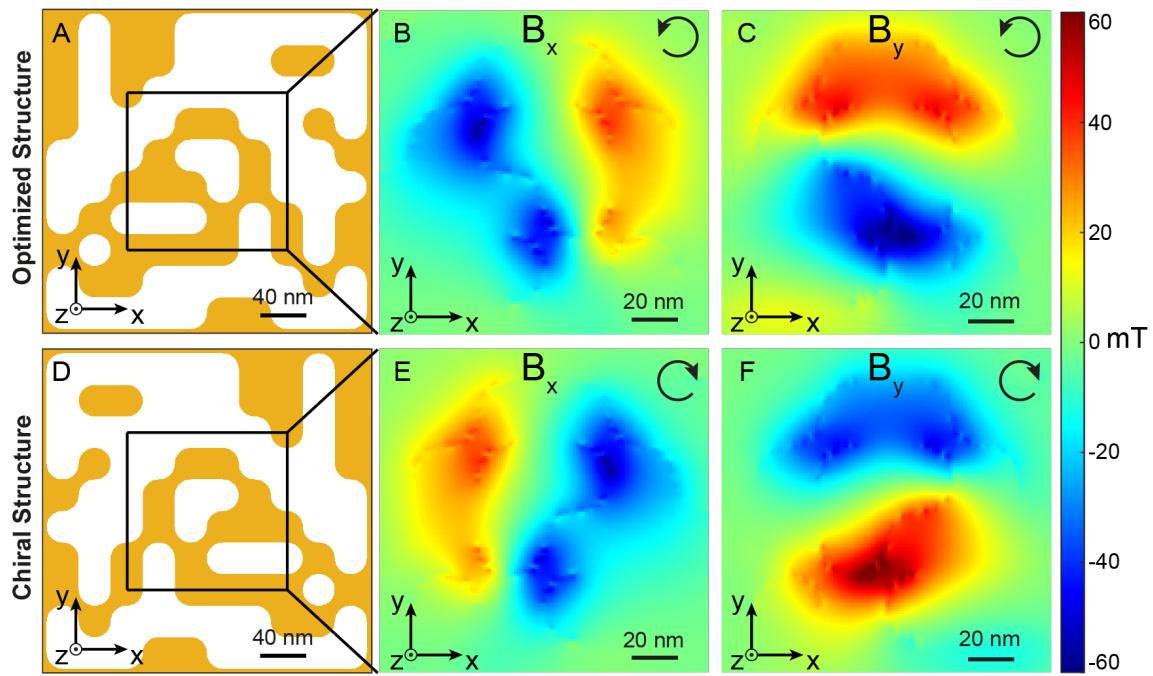

**Figure S4:** Distribution of drift currents in an XY plane. A) and B) Spatial distributions of drift currents at the surface of the optimized structure for the left and right circular polarizations of excitation, respectively. C) and D) Spatial distributions of drift currents at the surface of the mirror structure for the left and right circular polarizations of excitation, respectively.

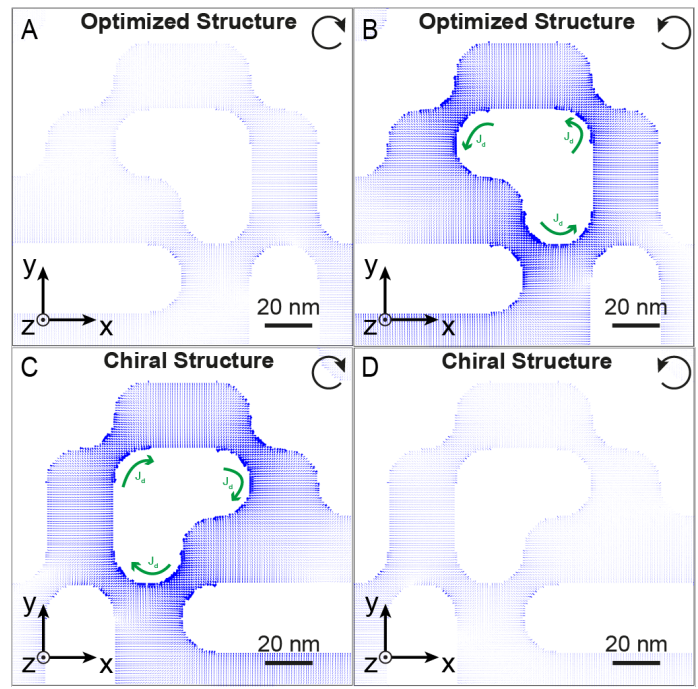

**Figure S5:** Magnetic field components. A) Schematic, in an XY plane, of the GA-optimized structure. B) and C) Spatial distributions of $B_x$ and $B_y$ generated at the Z-center of the structure shown in A) for the right circular polarizations of excitation. D) Schematic, in an XY plane, of the mirror structure. E) and F) Spatial distributions of $B_x$ and $B_y$ generated at the Z-center of the mirror structure shown in D) for the left circular polarizations of excitation.

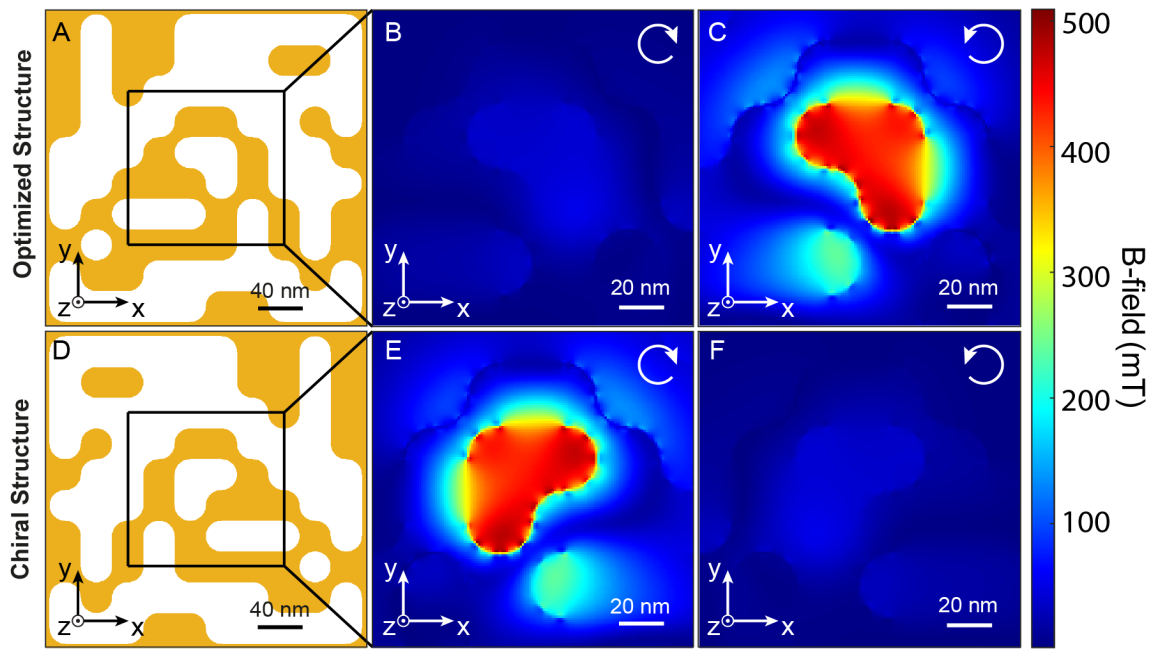

**Figure S6:** Amplitude of magnetic fields. A) Schematic, in an XY plane, of the GA-optimized structure. B) and C) Spatial distributions of the **B**-field amplitude at the Z-center of the structure shown in A) for the left and right circular polarizations of excitation, respectively. D) Schematic, in an XY plane, of the mirror structure. E) and F) Spatial distributions of **B**-field amplitude at the Z-center of the mirror structure shown in D) for the left and right circular polarizations of excitation, respectively.